\documentclass[aps,pre,reprint]{revtex4-1}
\pdfoutput=1

\usepackage[pdftex]{graphicx}
\usepackage{amsmath}
\usepackage{amsfonts}
\usepackage{mathrsfs}
\usepackage{bm}
\usepackage{hyperref}
\usepackage{times}

\begin{document}

\title{Partition function zeros of the $p$-state clock model in the complex temperature plane}

\author{Dong-Hee Kim}
\email{dongheekim@gist.ac.kr}
\affiliation{Department of Physics and Photon Science, School of Physics and Chemistry, Gwangju Institute of Science and Technology, Gwangju 61005, Korea}

\begin{abstract}
We investigate the partition function zeros of the two-dimensional $p$-state clock model in the complex temperature plane by using the Wang-Landau method. For $p=5$, $6$, $8$, and $10$, we propose a modified energy representation to enumerate exact irregular energy levels for the density of states without any binning artifacts. Comparing the leading zeros between different $p$'s, we provide strong evidence that the upper transition at $p=6$ is indeed of the Berezinskii-Kosterlitz-Thouless (BKT) type in contrast to the claim of the previous Fisher zero study [Phys. Rev. E \textbf{80}, 042103 (2009)]. We find that the leading zeros of $p=6$ at the upper transition collapse onto the zero trajectories of the larger $p$'s including the $XY$ limit while the finite-size behavior of $p=5$ differs from the converged behavior of $p \ge 6$ within the system sizes examined. In addition, we argue that the nondivergent specific heat in the BKT transition is responsible for the small partition function magnitude that decreases exponentially with increasing system size near the leading zero, fundamentally limiting access to large systems in search for zeros with an estimator under finite statistical fluctuations.
\end{abstract}

\maketitle

\section{Introduction}

The Berezinskii-Kosterlitz-Thouless (BKT) transition~\cite{Berezinskii1971,KT1972} has attracted steady attention because of its physical richness and generality in explaining the stabilization of quasi-long-range order in two-dimensional (2D) systems with continuous symmetry~\cite{BKTreview}. The classical 2D $XY$ model is probably the most extensively studied example showing the BKT transition, often being used as a reference of its peculiar critical behavior at the transition point and universal features~\cite{BKTreview,KT1973,KT1974,Jose1977,Kenna2005}. While continuous symmetry is essential for the BKT transitions, it can also emerge from a system without explicit continuous symmetry. The $p$-state clock model is a cousin of the $XY$ model with discrete $\mathrm{Z}_p$ symmetry. The Hamiltonian of the clock model is written as
\begin{equation}
\mathcal{H} = -J\sum_{\langle i,j \rangle} \cos (\theta_i^{(p)} - \theta_j^{(p)} ),
\end{equation}
where $J>0$ is the ferromagnetic coupling given between a nearest-neighbor pair of spins with discrete angle variables $\theta^{(p)} = 2\pi n / p$ for $n \in \{0,\ldots,p-1\}$. While the exact $XY$ model is recovered only in the limit of infinite $p$, it was found that the BKT characters would appear in the $\mathrm{Z}_p$ models when $p \gtrsim 5$~\cite{Elitzur1979,Cardy1980,Frohlich1981,Ortiz2012}. The nature of phase transitions in the general clock model has been widely studied with different theoretical and numerical approaches, which, however in some parts, have given mixed results on the characterization of transitions around the lower bound of $p$ (for instance, see the summary of the related debates in Ref.~\cite{Borisenko2011}).

The Villain formulation of the $\mathrm{Z}_p$ model showed that when $p > 4$, the phase diagram consists of three different areas where the intermediate massless phase undergoes two BKT transitions into the high-temperature disordered and low-temperature ordered phases~\cite{Elitzur1979,Einhorn1980,Hamer1980,Nienhuis1984}. In the standard clock model, the Monte Carlo (MC) simulations with the phenomenological finite-size-scaling analysis~\cite{Tobochnik1982,Challa1986,Tomita2002,Borisenko2011} indeed found the critical exponents for $p \ge 5$ that are consistent with the theoretical predictions~\cite{Jose1977,Elitzur1979}. On the other hand, differences from the BKT transition of the $XY$ limit have also been argued in the studies of different measures. At $p=5$, it was observed that the helicity modulus does not vanish in the disordered phase~\cite{Baek2010b,Baek2013}, which disagrees with the universal jump from zero expected in the BKT transition~\cite{Nelson1977,Minnhagen1981} and observed in the systems of $p=6$~\cite{Baek2010a} and above~\cite{Lapilli2006}. Later, the helicity modulus redefined with a finite twist matching the discrete symmetry resolved this issue~\cite{Kumano2013}, providing consistent estimates of the transition temperatures~\cite{Kumano2013,Chatelain2014}. 

At $p=6$, the disagreement that remains unresolved is with the previous scaling tests of the leading Fisher zeros of the partition function claiming that the transitions in the six-state clock model may not be of BKT type~\cite{Hwang2009}. While this claim supported the earlier test of the helicity modulus~\cite{Lapilli2006}, the later calculations of the helicity modulus in larger systems agreed on the existence of the BKT transitions at $p=6$~\cite{Baek2010a,Kumano2013,Baek2013,Chatelain2014}. However, the Fisher zero issue raised at $p=6$ remains unexamined so far, and moreover there has been no Fisher zero study attempted for other $p$'s at all. In this paper, we report the first comparative calculation of the leading Fisher zeros for $p=5$, $6$, $8$, and $10$.

The main question that we address here is how the leading Fisher zeros evolve with increasing $p$ and more specifically how different the zeros of $p=6$ are from those of large $p$'s that are known to exhibit the BKT transitions. We perform extensive numerical calculations based on the Wang-Landau (WL) sampling of the density of states (DOS). We find that at the upper transition, the leading zeros of $p=6$ are in fact collapsed onto the trajectory of the larger $p$'s including the $XY$ limit, providing strong evidence that the transition at $p=6$ is indeed of BKT type in contrast to the claim based on the previous scaling tests within the six-state clock model~\cite{Hwang2009}. 

For the limited system sizes that are accessible in numerically finding the Fisher zero within the WL DOS samples, finite-size corrections naturally affect the analysis at the level of an individual $p$, which is apparent in the previous test at $p=6$~\cite{Hwang2009} and in our observation of the distinguished finite-size behavior at $p=5$. Remarkably, the collapsed Fisher zero trajectory that we observe for $p \ge 6$ indicates that the finite-size effect becomes also well converged between different $p$'s when $p \ge 6$, demonstrating the advantage of the comparative approach that allows us to infer the transition class of $p=6$ deductively from the known BKT character of the larger $p$'s.

On the numerical side, we provide a modified representation of the Hamiltonian for the considered group of $p$'s that enables exact energy enumeration, which is crucial to our application of the WL method~\cite{WL1,WL2} to the Fisher zero problem in the $p$-state clock model. The usual WL approach benefits from regularly spaced energy levels, which, however, is not the case in the cosine energy of the clock model except for the very special case of $p=6$. Here we find that for a group of $p$'s, the irregular energy structure can be decomposed into two regular parts, allowing full energy resolution in building the DOS by using the 2D WL procedures without any necessity of introducing artificially binned energy space. 

This paper is organized as follows. Section~\ref{sec:method} describes our method of the exact energy enumeration and the details of the WL procedures. The two-step method of the Fisher zero finder is also briefly explained. In Sec.~\ref{sec:result}, we present our main results of a comparison between the leading zeros computed for $p=5$, $6$, $8$, and $10$. The implications of the collapsed leading zero trajectories that are found for $p\ge 6$ are discussed. An analysis of numerical uncertainty is also given in this section, and the connection with the specific heat at the BKT transition is argued. Finally, conclusions are given in Sec.~\ref{sec:conclusion}. 

\section{Numerical methods}
\label{sec:method}

The connection between the singular behavior of free energy and the zeros of the partition function was first formulated by Yang and Lee in the plane of complex fugacity~\cite{YangLeeZero}, and then the Fisher zero that we focus on here was proposed for a canonical partition function in complex temperature~\cite{FisherZero}. Their usefulness has been demonstrated in various model systems and was recently also emphasized by experimental observations~\cite{Peng2015,Brandner2016}. Although the behavior of the leading zeros closest to the real axis is well established in the second- and first-order phase transitions  (see, for instance, Ref.~\cite{Janke2001} and references therein), it has been extended to the BKT transition only very recently with the $XY$ model by using the higher-order tensor renormalization-group (HOTRG)~\cite{Denbleyker2014} and the WL method with energy binning~\cite{Rocha2016,Costa2017}. 

In this section, we present our extension of the WL method to the leading zero calculations for the $p$-state clock models, which is designed to avoid the energy binning.  

\subsection{Wang-Landau formulation of the $p$-state clock model}

\begin{table}
\caption{Two-term representation of the Hamiltonian for $p=5$, $6$, $8$, and $10$. The index $n$ represents the possible values of $|n_i - n_j|$ where the spin angle variable $n_{i,j} \in \{0,\ldots,p-1\}$.}
\begin{ruledtabular}
\begin{tabular}{crrrrrrrrrr|c}
$n$      & $0$ &  $1$ &  $2$ &  $3$ &  $4$ &  $5$ &  $6$ &  $7$ &  $8$ & $9$ & $J^{(1)}_p/J$ \\ \hline
$\mathcal{E}^{(1)}_5(n)$    & $4$ & $-1$ & $-1$ & $-1$ & $-1$ &      &      &      &      &     & $1/4$ \\
$\mathcal{E}^{(1)}_6(n)$    & $2$ & $1$  & $-1$ & $-2$ & $-1$ & $1$  &      &      &      &     & $1/2$ \\ 
$\mathcal{E}^{(1)}_8(n)$    & $1$ &  $0$ &  $0$ &  $0$ & $-1$ &  $0$ &  $0$ &  $0$ &      &     & $1$   \\
$\mathcal{E}^{(1)}_{10}(n)$ & $4$ &  $1$ & $-1$ &  $1$ & $-1$ & $-4$ & $-1$ &  $1$ & $-1$ & $1$ & $1/4$ \\ 
\hline \hline
$n$      & $0$ &  $1$ &  $2$ &  $3$ &  $4$ &  $5$ &  $6$ &  $7$ &  $8$ & $9$ & $J^{(2)}_p/J$ \\ \hline
$\mathcal{E}^{(2)}_5(n)$    & $0$ &  $1$ & $-1$ & $-1$ &  $1$ &      &      &      &      &     & $\sqrt{5}/4$ \\
$\mathcal{E}^{(2)}_6(n)$    & $0$ &  $0$ &  $0$ & $0$  &  $0$ & $0$  &      &      &      &     & $0$ \\
$\mathcal{E}^{(2)}_8(n)$    & $0$ &  $1$ &  $0$ & $-1$ &  $0$ & $-1$ &  $0$ &  $1$ &      &     & $1/\sqrt{2}$ \\
$\mathcal{E}^{(2)}_{10}(n)$ & $0$ &  $1$ &  $1$ & $-1$ & $-1$ &  $0$ & $-1$ & $-1$ &  $1$ & $1$ & $\sqrt{5}/4$ \\ 
\end{tabular}
\end{ruledtabular}
\label{tab:mixedH}
\end{table}

While the WL method in conjunction with a polynomial solver has often been used to calculate the Fisher zeros in spin models~\cite{Rocha2016,Costa2017,Rocha2014,Taylor2013,Lee2010}, it cannot be directly applied to a general $p$-state clock model. Irregularly spaced energies from the sum of cosines in the clock model cause a large numerical challenge in the WL sampling, and a polynomial expansion of the partition function is simply not possible with this exact energy structure being kept. Note that the previous case of $p=6$~\cite{Hwang2009} is an exception since its energy is given as an integer-multiple of $J/2$. Probably the easiest way to deal with the irregularity is to introduce an extra energy binning step, which, however, comes with an unavoidable loss of spectral resolution.  

Nevertheless, we find that for a group of $p$'s including $5$, $8$, and $10$, the energies can be mapped onto the two-dimensional regular grids where the dimensions represent the rational and irrational parts of the cosine energy~\cite{footnote1}. The Hamiltonian is accordingly decomposed into two terms as
\begin{equation}
\mathcal{H} = -J^{(1)}_p \sum_{\langle i,j \rangle} \mathcal{E}^{(1)}_p(n_{ij}) - J^{(2)}_p \sum_{\langle i,j \rangle} \mathcal{E}^{(2)}_p(n_{ij}),
\label{eq:newH}
\end{equation} 
where $n_{ij} \equiv |n_i - n_j|$ is the spin angle difference. The functions $\mathcal{E}^{(1)}_p$ and $\mathcal{E}^{(2)}_p$ are integer-valued as tabulated in Table~\ref{tab:mixedH}. Therefore, for such $p$'s, one finds $\mathcal{H} \equiv \mathcal{H}(E_1,E_2) = -J^{(1)}_p E_1 - J^{(2)}_p E_2$ being represented by two integers of $E_1 \equiv  \sum_{\langle i,j \rangle} \mathcal{E}^{(1)}_p$ and $E_2 \equiv \sum_{\langle i,j \rangle} \mathcal{E}^{(2)}_p$, which allows efficient numerics using a standard array for random walks in energy space without loss of precision. 

The joint DOS $g(E_1,E_2)$ for the combinations of $E_1$ and $E_2$ is then evaluated by the WL sampling through the 2D random walk processes~\cite{Landau2004, Zhou2006, Silva2006, Tsai2007, Kwak2015}. Although the increased dimensionality requires a long computational time in exchange for having an exact access to the energy levels, our implementation handles about three million energy levels in the largest calculation performed for $L=20$ at $p=10$. The system size is denoted by $L$ representing $L^2$ sites of our square lattices. In the WL procedures, we follow the standard strategy to decrease the modification factor (see, for instance, Ref.~\cite{Kwak2015}). We set the histogram flatness criterion to be $0.99$ for all $p=6$ cases and for small systems of other $p$'s; it is lowered to $0.95$ when $L > 12$ for $p=5$ and $8$; for $p=10$, it is $0.95$ when $8 < L < 16$ and $0.9$ when $L$ is larger. We obtain $30$ samples of the WL DOS from independent runs at each $p$ to evaluate the uncertainty of estimates through a resampling process. 

\subsection{Partition function zero calculations}

Since the WL method provides unnormalized samples of the DOS, we consider the normalized partition function $\tilde{\mathcal{Z}}(\beta)$ in complex inverse temperature $\beta \equiv \beta_R + i \beta_I$, defined as 
\begin{equation}
\tilde{\mathcal{Z}}(\beta)\equiv\frac{\mathcal{Z}(\beta)}{\mathcal{Z}(\beta_R)} = \sum_{E_1,E_2} P(E_1,E_2;\beta_R) e^{-i\beta_I \mathcal{H}},
\end{equation}
where the energy distribution at a real temperature $\beta_R$ is
\begin{equation}
P(E_1,E_2;\beta_R) \equiv \frac{1}{\mathcal{Z}(\beta_R)}g(E_1,E_2)e^{-\beta_R \mathcal{H}(E_1,E_2)}.
\end{equation}
The partition function $\mathcal{Z}(\beta_R)\equiv \sum g \exp({-\beta_R\mathcal{H}})$ at a real temperature $\beta_R$ is nonzero in a finite system. An arbitrary normalization of a WL DOS sample $g(E_1,E_2)$ is then canceled out, and thus it has no effect on the energy distribution and the normalized partition function. Using multiple WL samples of $g(E_1,E_2)$, we replace $P(E_1,E_2;\beta_R)$ with the sample-averaged one $\langle P(E_1,E_2;\beta_R) \rangle_\mathrm{WL}$. The uncertainty is estimated with respect to this average over the WL samples for a 95\% confidence interval from the bootstrap resampling processes repeated for $1000$ times. 

Once the WL samples of DOS $g(E_1,E_2)$ are obtained, one can compute the normalized partition function for any given complex temperature without restriction, which is a numerical advantage of the WL method over the histogram reweighting MC calculations. Since the polynomial expansion is not simple with two variables, the complex plane of $\beta$ is searched for the zeros of the partition function by using the two-step method~\cite{Falcioni1982, Alves1992, Denbleyker2014}. 

For a given $\beta_R$, the real and imaginary parts of $\tilde{\mathcal{Z}}$ are smooth oscillating functions of $\beta_I$, and thus a set of the zeros in the axis of $\beta_I$ can be easily found for each oscillation, constructing a map of the zeros of $\mathrm{Re}[\tilde{\mathcal{Z}}]$ and $\mathrm{Im}[\tilde{\mathcal{Z}}]$ in the complex $\beta$ plane. First, an intersection point between the zero curves of $\mathrm{Re}[\tilde{\mathcal{Z}}]$ and $\mathrm{Im}[\tilde{\mathcal{Z}}]$ on this map is graphically located. Second, the function $|\tilde{\mathcal{Z}}|^2$ is numerically minimized around the graphical intersection to precisely locate the zero of $\tilde{\mathcal{Z}}(\beta)$. Through these steps, the leading zero $\beta_1$ with the smallest imaginary part is identified in each area of the upper and lower transitions~\cite{SM}.

\section{Results and Discussions}
\label{sec:result}

\begin{figure}
\includegraphics[width=0.47\textwidth]{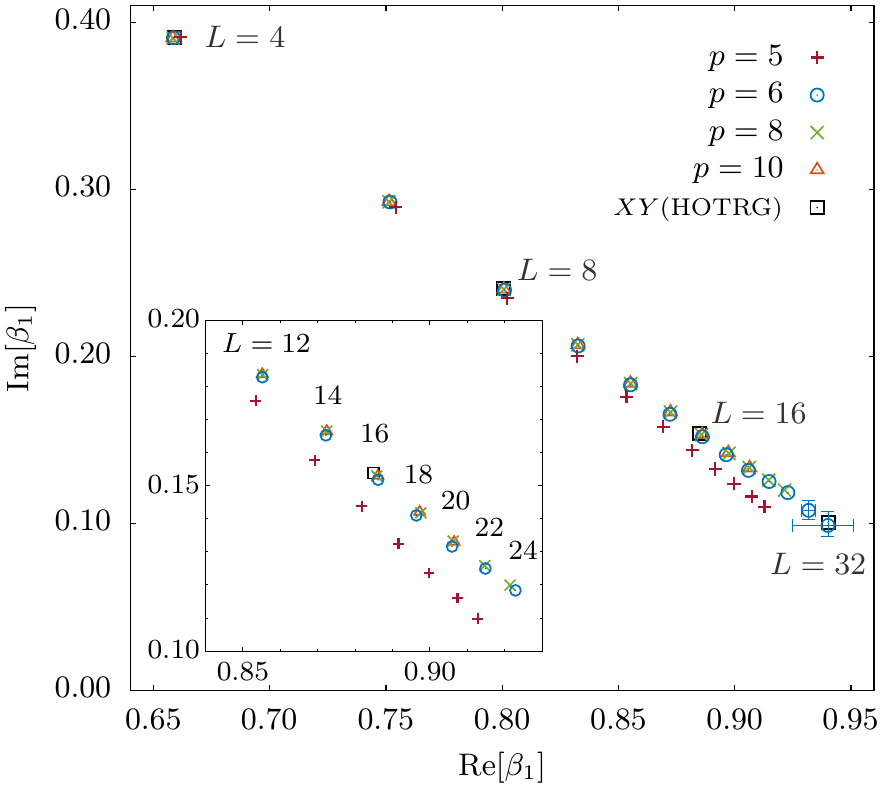}
\caption{Leading Fisher zeros at the upper transition in the $p$-state clock model with $p=5$, $6$, $8$, and $10$. The uncertainty shown by the error bar is given as a 95\% confidence interval estimated from the bootstrap resampling with the WL DOS samples. The error bar is omitted if it is smaller than the symbol size. The data points for the $XY$ limit are from the previous higher-order tensor renormalization-group (HOTRG) calculations~\cite{Denbleyker2014}. }
\label{fig1}
\end{figure}

Figure~\ref{fig1} displays the leading Fisher zeros identified at the upper transition area in the $p$-state clock models of $p=5$, $6$, $8$, and $10$. We find that the calculated leading zeros of $p \ge 6$ collectively move in the complex temperature plane. We also compare the leading zeros of finite $p$'s with the data points of the 2D $XY$ model that are available in the previous higher-order tensor renormalization-group (HOTRG) calculations~\cite{Denbleyker2014}. Notably, it turns out that for $p\ge 6$, the locations of the zeros become well collapsed onto the leading zeros of the $XY$ model. The converged trajectory observed at $p \ge 6$ strongly suggests that the upper transition at $p=6$ indeed belongs to the same BKT transition of the $XY$ model.

This is in clear contrast to the claim in the previous Fisher zero study of the six-state clock model~\cite{Hwang2009}, which argued that the transitions at $p=6$ may not be of BKT type. The previous work was based on the finite-size-scaling analysis on the leading zeros that actually fitted well into either ansatz of the BKT or second-order transitions. Our approach is different in the following sense. Instead of trying to distinguish the order of a transition based on the finite-size-scaling analysis on a model of an individual $p$, we compare the leading-zero trajectories between different $p$'s to find their converged behavior. Given that the common nature of their BKT transitions at $p=8$ and $10$ and in the $XY$ model is well established, the observed convergence can lead us to infer that the model of $p=6$ is in the same class of the larger $p$'s. 

\begin{figure}
\includegraphics[width=0.47\textwidth]{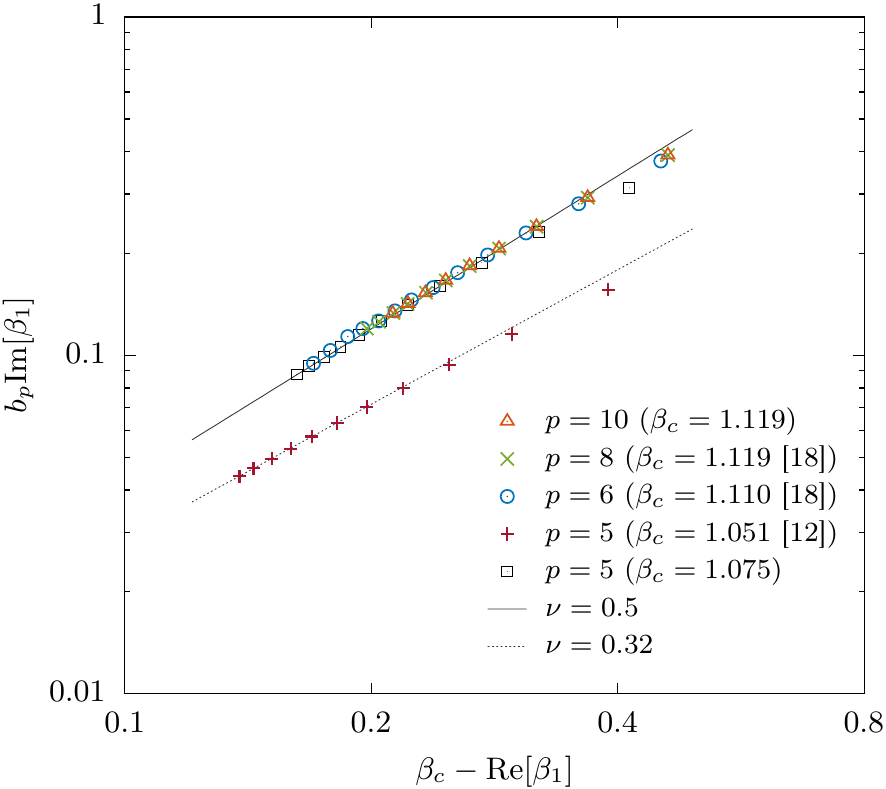}
\caption{Scaling relation between the real and imaginary parts of the leading Fisher zero. The power-law relation $\mathrm{Im}[\beta_1]\propto (\beta_c - \mathrm{Re}[\beta_1])^{1+\nu}$ is examined with the previous estimates of critical points $\beta_c$ \cite{Borisenko2011,Tomita2002}. The arbitrary factor $b_p$ is adjusted for a graphical comparison between the data points of different $p$'s.}
\label{fig2}
\end{figure}

The same BKT character of $p\ge 6$ is supported by the mutual collapse of their leading-zero trajectories onto a common power-law curve shifted by the known transition points. Extending the finite-size-scaling ansatz of the correlation length to the complex temperature domain, the analysis for the $XY$ model~\cite{Denbleyker2014} suggested that the leading zero moves toward the real axis along the power-law trajectory,
\begin{equation}
\beta_I \propto (\beta_c - \beta_R)^{1+\nu},
\end{equation}   
in the area of small $\beta_I$. In Fig.~\ref{fig2}, we examine this power-law relation for the common BKT exponent $\nu=0.5$ by using the transition temperatures provided by the previous MC results. We find that the upper transition points of $\beta_c \approx 1.110$ for $p=6$ \cite{Tomita2002} (see also \cite{Baek2010a, Kumano2013}) and $\beta_c \approx 1.119$ for $p\ge 8$ \cite{Tomita2002} lead to good collapse of the data points falling onto the power-law curve with exponent $\nu=0.5$. 

The universal behavior observed for $p\ge 6$ implies that their finite-size influences are also indistinguishable between those $p$'s. This fast convergence of the finite-size effects is remarkable considering the limited accessible system sizes in our calculations. Although it is natural to anticipate that the finite-size corrections play a role in such small systems, the collapse of the leading-zero trajectories suggests that the finite-size effect becomes nearly independent of $p$ when $p \ge 6$.   

On the other hand, the leading-zero trajectory of $p=5$ shows an apparent deviation from those of the larger $p$'s, which indicates a very different type of finite-size effects appearing in its transition point and the scaling exponent. With the transition point being fixed at the previous MC estimate of $\beta_c \simeq 1.051$~\cite{Borisenko2011}, the leading-zero trajectory of $p=5$ does not fall onto the curve with $\nu=0.5$, giving a better fit to the one with $\nu \simeq 0.32$ within the system sizes that are accessible. The other estimates from the helicity modulus with finite twist, $\beta_c \simeq 1.059$~\cite{Kumano2013} and $1.058$~\cite{Chatelain2014}, provide a larger value of $\nu \simeq 0.38$. Adjusting a transition point to be $1.075$ causes the curve to get closer to the one with $\nu=0.5$, but the curve still deviates from the line of the larger $p$'s. 

While these strong finite effects at $p=5$ are distinguished from the well-converged behavior in the trajectories of the larger $p$'s, this deviation should not be misinterpreted as evidence of a different transition nature. Indeed, a strong finite-size effect at $p=5$ has also been witnessed with a different measure. In the previous study of the helicity modulus with a finite twist, the finite-size behavior of the helicity modulus was indicated at $p=5$ in the intermediate BKT region, while at $p=6$, it was almost independent of the system size as predicted in the BKT phase~\cite{Kumano2013}.     

\begin{figure}
\includegraphics[width=0.47\textwidth]{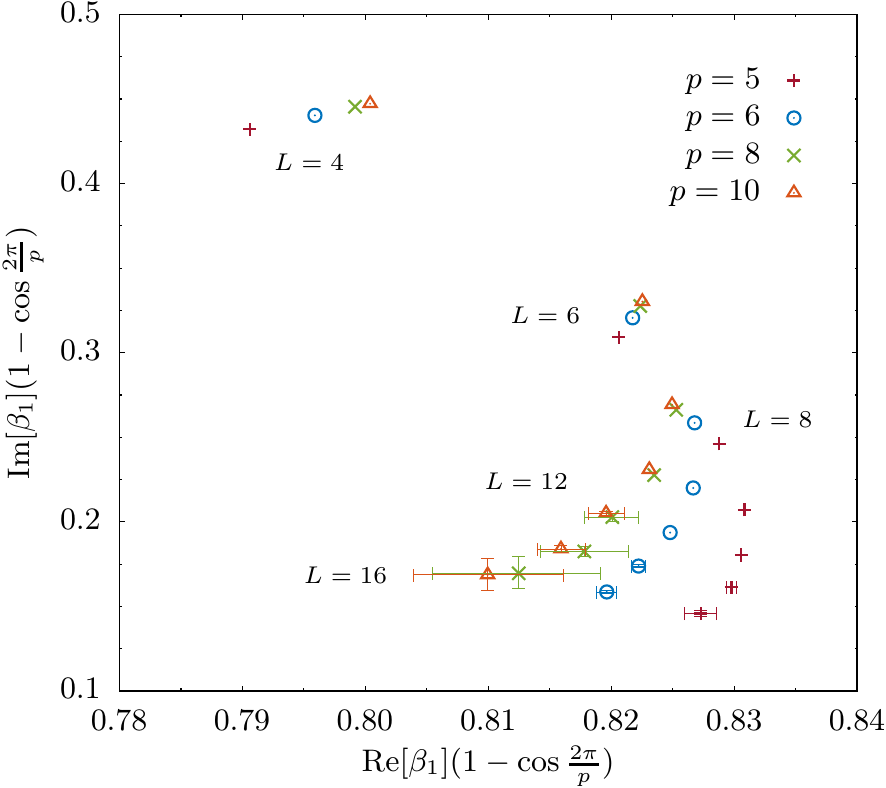}
\caption{Leading Fisher zeros at the lower transition. The real and imaginary parts of the leading zeros are rescaled with factor $(1-\cos\frac{2\pi}{p})$. The system sizes are limited to $L \le 16$ for reliable identification of the leading zeros.}
\label{fig3}
\end{figure}

In addition, we also calculate the leading zeros in the lower-temperature side of the two transitions. Figure~\ref{fig3} presents the $p$ dependence of the leading zeros with rescaling. We show that the trajectory of the corresponding leading zeros moves systematically toward the zero-temperature limit of the complex $\beta$ plane as $p$ increases. In the Peierls argument~\cite{Ortiz2012}, the transition temperature would scale as $T_c \sim (1-\cos\frac{2\pi}{p})$, which recovers the $1/p^2$ behavior in the limit of large $p$. For the leading zeros, we find that both the real and imaginary parts of the zeros scale roughly with the same factor $1/(1-\cos\frac{2\pi}{p})$, showing a trend in which the trajectory of the leading zeros approaches a common curve as $p$ increases.

While the converged trajectory of the leading zeros that we have found for $p\ge 6$ at the upper transition is already clear within the system sizes examined, it is still important to precisely know the numerical limitations encountered when simulating larger systems. This would clarify the challenge in performing a conventional finite-size-scaling analysis of an explicit system-size dependence, which is avoided in our present study. For instance, it is expected that the imaginary part of the leading zero scales with system size $L$ as $\beta_I \sim (\ln bL)^{-\tilde{q}}$, where $\tilde{q} = 1 + 1/\nu$ for small $\beta_I$ in the BKT transition~\cite{Denbleyker2014}. Comparing such a logarithmic form with the power-law ansatz of the second-order transition would hardly be conclusive in small systems as was already noticed in the previous Fisher zero study of the six-state clock model~\cite{Hwang2009}. 

The numerical bottleneck is twofold in our calculations. The obvious one is the well-known large cost in computational time required for the 2D WL procedures that are essential for $p=5$, $8$, and $10$. It is hard in practice to go beyond a system of a few million energy levels. This might be improved in the future by a proposed extension of the parallel WL algorithm~\cite{Vogel2013,Vogel2014} to 2D energy space~\cite{Valentim2015,Ren2016,Chan2017}. In addition, the special 1D WL case of $p=6$ does not suffer from a such problem since the number of energy levels scales linearly with the number of lattice sites. We have been able to reach easily up to $L=128$ in the case of $p=6$.  

The more critical issue is the explosively growing uncertainty in locating the leading Fisher zeros as the system size increases. This can be best seen in the larger-system calculations at $p=6$ where a sudden increase of the uncertainty occurs at $L=32$ (see Fig.~\ref{fig1}) and is generally observed in all calculations that we have done. While the only source of the errors in our numerics is the stochastic WL process itself, below we explain how the small stochastic noises can be amplified quickly in the Fisher zero calculations for the $p$-state clock model and its fundamental connection to the BKT transition. 

\begin{figure}
\includegraphics[width=0.47\textwidth]{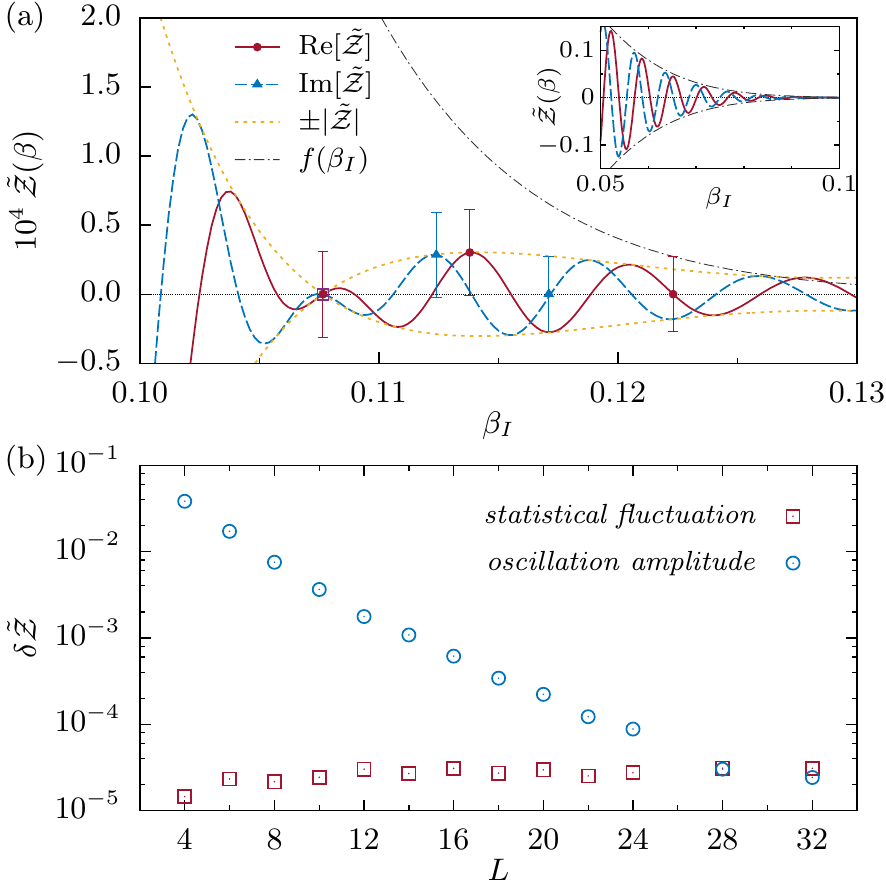}
\caption{Reliability test of the leading-zero identification for the upper transition in the six-state clock model. (a) The normalized partition function $\tilde{\mathcal{Z}}(\beta)$ is evaluated for $L=28$ as a function of $\beta_I$ at $\beta_R = \mathrm{Re}[\beta_1]$. The leading zero is marked by the square symbol. The error bar displayed at the points along the oscillations presents the statistical fluctuation measured by bootstrap resampling, indicating that its magnitude is typical in the range of $\beta_I$ for both of $\mathrm{Re}[\tilde{\mathcal{Z}}]$ and $\mathrm{Im}[\tilde{\mathcal{Z}}]$. The envelope function $f(\beta_I)$ is obtained from the Gaussian approximation~\cite{Alves1992}. (b) The maximum oscillation amplitude of $\mathrm{Re}[\tilde{\mathcal{Z}}]$ for $\beta_I > \mathrm{Im}[\beta_1]$ is shown for comparison with its fluctuation as a function of system size $L$.}
\label{fig4}
\end{figure}

Figure~\ref{fig4} demonstrates how the uncertainty develops in finding the leading zero at $p=6$, where the WL simulations can be done for relatively large systems while maintaining the accuracy of the DOS samples at the same high level. In the system of $L=28$ shown in Fig.~\ref{fig4}(a), the fluctuation of the partition function $\mathcal{\tilde{Z}}$ turns out to be almost comparable to the maximum oscillation amplitude in the region of $\beta_I > \mathrm{Im}[\beta_1]$. This implies that for smaller oscillation amplitude, the oscillatory behavior could be completely buried in the scale of the fluctuation, making our zero search unreliable. Therefore, the accuracy of the zero identified is guaranteed only when the WL estimate of $\mathcal{\tilde{Z}}$ has an oscillation amplitude larger than its statistical fluctuation in the vicinity of the zero.  

We find that in the $p$-state clock model, the oscillation amplitude of $\tilde{\mathcal{Z}}$ near the leading zero decreases exponentially with increasing system size $L$, as shown in Fig.~\ref{fig4}(b) for the case of $p=6$. The zero search in this case undergoes a crossover around $L=28$ above which the fluctuation gets larger than the oscillation amplitude. This implies that considering a larger system for proper finite-size-scaling analysis would require extreme accuracy of a DOS estimate to cope with the exponentially decreasing oscillation amplitude of $\tilde{\mathcal{Z}}$. In the $p$-state clock model that we consider, this can be an important issue for the Fisher zero search within the MC methods that essentially come with statistical noises.

The exponential system-size scaling of $\tilde{\mathcal{Z}}$ and the resulting tight bound of the accessible system size is perhaps a consequence of the BKT transition where the specific heat is nondivergent~\cite{Tobochnik1982,Kenna2005,Borisenko2011}. In the Gaussian approximation of energy distribution~\cite{Alves1992}, at a given $\beta_R$, the envelope function of $\tilde{\mathcal{Z}}(\beta_I)$ is calculated as $f(\beta_I)=\exp [ - C\beta_I^2 / 2\beta_R^2 ]$, where $C$ denotes heat capacity at $\beta_R$. While the Gaussian approximation is not valid at the zero, it may still work as an upper bound of the oscillation amplitudes in its vicinity, as indicated in Fig.~\ref{fig4}(a). From the scaling forms $\beta_I \sim (\ln bL)^{-\tilde{q}}$ and $C \sim L^2$, one can see that $f(\beta_I)$ behaves as $\exp [ - a L^2 (\ln bL)^{-2\tilde{q}}]$ near the zero, which provides a rough sketch of the extreme accuracy requirement to increase the system size. 

\begin{table*}
\caption{The leading zeros from the Wang-Landau simulations of the $p$-state clock model at the upper transition.}
\begin{ruledtabular}
\begin{tabular}{ccccc}
$L$  & $p=5$ & $p=6$ & $p=8$ & $p=10$\\ \hline
$4$  & $0.66172011+0.39093662i$ & $0.65860056+0.39064303i$ & $0.65840832+0.39041109i$ & $0.65837736+0.39042101i$\\
$6$  & $0.75420037+0.28919617i$ & $0.75171500+0.29228589i$ & $0.75131882+0.29245805i$ & $0.75132177+0.29244077i$\\
$8$  & $0.80211731+0.23458419i$ & $0.80102406+0.23961812i$ & $0.80071947+0.24008289i$ & $0.80071719+0.24003913i$\\
$10$ & $0.83229831+0.19990540i$ & $0.83266310+0.20612369i$ & $0.83248717+0.20686393i$ & $0.83254841+0.20681163i$\\
$12$ & $0.85348475+0.17570736i$ & $0.85520368+0.18283498i$ & $0.85528976+0.18355279i$ & $0.85517823+0.18365398i$\\
$14$ & $0.86925922+0.15773549i$ & $0.87217137+0.16522828i$ & $0.87240221+0.16648854i$ & $0.87243520+0.16642103i$\\
$16$ & $0.88179311+0.14381439i$ & $0.88615585+0.15177668i$ & $0.88579947+0.15300988i$ & $0.88605832+0.15290659i$\\
$18$ & $0.89159755+0.13243507i$ & $0.89639149+0.14099277i$ & $0.89758758+0.14180194i$ & $0.89724782+0.14204027i$\\
$20$ & $0.89976717+0.12358720i$ & $0.90594030+0.13159544i$ & $0.90628488+0.13322776i$ & $0.90647873+0.13295482i$\\
$22$ & $0.90739307+0.11603649i$ & $0.91484763+0.12496506i$ & $0.91466498+0.12579545i$ &\\
$24$ & $0.91287777+0.10978935i$ & $0.92291258+0.11833457i$ & $0.92145896+0.11991071i$ &\\
$28$ &                          & $0.93185468+0.10765101i$ & &\\
$32$ &                          & $0.94010329+0.09861775i$ & &
\end{tabular}
\end{ruledtabular}
\label{tab:zero_upper}
\end{table*}

\begin{table*}
\caption{The leading zeros from the Wang-Landau simulations of of the $p$-state clock model at the lower transition.}
\begin{ruledtabular}
\begin{tabular}{ccccc}
$L$  & $p=5$ & $p=6$ & $p=8$ & $p=10$\\ \hline
$4$  & $1.14415021+0.62547361i$ & $1.59178785+0.88047270i$ & $2.72848984+1.52074963i$ & $4.19093387+2.34143852i$\\
$6$  & $1.18762295+0.44754788i$ & $1.64349226+0.64124630i$ & $2.80774534+1.11857803i$ & $4.30682954+1.72881228i$\\
$8$  & $1.19942768+0.35623721i$ & $1.65356403+0.51705887i$ & $2.81771511+0.90898000i$ & $4.31949110+1.40970891i$\\
$10$ & $1.20239494+0.29971424i$ & $1.65333971+0.44008450i$ & $2.81156339+0.77716843i$ & $4.30980838+1.20851334i$\\
$12$ & $1.20203326+0.26117339i$ & $1.64959572+0.38737205i$ & $2.79998220+0.69237318i$ & $4.29136359+1.07360761i$\\
$14$ & $1.20086668+0.23365741i$ & $1.64443723+0.34782938i$ & $2.79219279+0.62308855i$ & $4.27214959+0.96335725i$\\
$16$ & $1.19725528+0.21117272i$ & $1.63927235+0.31715739i$ & $2.77399083+0.57881273i$ & $4.24092498+0.88298542i$
\end{tabular}
\end{ruledtabular}
\label{tab:zero_lower}
\end{table*}

\section{Conclusions and Remarks}
\label{sec:conclusion}

We have investigated the leading Fisher zeros of the $p$-state clock model in square lattices by introducing the Wang-Landau formulation with exact energy enumeration for $p=5$, $6$, $8$, and $10$. We have found that the leading Fisher zeros show a converged trajectory at the upper transition when $p \ge 6$ including the $XY$ limit, providing strong evidence that the model with $p=6$ is in the same class with the larger $p$'s exhibiting the BKT transition. This is in contrast to the claim of the previous Fisher zero study for the six-state clock model~\cite{Hwang2009}, which argued that the transitions may not be of BKT type. Indeed, our findings are consistent with all up-to-date helicity modulus calculations~\cite{Baek2010a,Kumano2013,Baek2013,Chatelain2014}, which would help to resolve the remaining discrepancy between different numerical approaches characterizing the transitions in the six-state clock model in two dimensions.

It is also interesting to see a possibility that the converged behavior with increasing $p$ could be a general feature of the $p$-state clock models in different settings~\cite{Lupo}. For instance, it was recently reported that in the spin glass $p$-state clock model on diluted graphs, a physical observable converges quickly to the $XY$ limit as $p$ increases~\cite{Lupo2017}. The spin glass clock models on different underlying geometries have been argued to be indeed in the same class of their $XY$ limits when $p\gtrsim 5$~\cite{Nobre1986,Ilker2013,Ilker2014}, suggesting a very similar role of the discrete symmetry existing in general clock models.  

We have also argued that the numerical accessibility to the leading zero is closely related to the characteristic specific heat at a phase transition. For a divergent specific heat at the first- or second-order transition, the decreasing behavior of the imaginary part of the zero $\beta_I$ is canceled out by the divergence of the heat capacity. In the case of the first-order transition in $d$ dimensions, the factor $\exp[-C\beta_I^2]$ becomes $\mathcal{O}(1)$ since heat capacity $C \sim L^{2d}$ while $\beta_I \sim L^{-d}$; in the second-order transition, the scaling forms $C \sim L^{\alpha/\nu+d}$ and $\beta_I \sim L^{-1/\nu}$ provide the same result through the hyperscaling relation when $\alpha > 0$. The system-size dependence of the uncertainty in the zero finder could be further quantifiable by the confidence range for $\beta_I$~\cite{Alves1992}. Although it is necessary to examine this expectation numerically in real models as it is based on the Gaussian approximation, it raises a possibility that the range of system sizes accessible with estimates under statistical noises is much wider for the ordinary phase transitions than for the BKT transitions in the search for Fisher zeros.
 
\begin{acknowledgments}
We thanks Chi-Ok Hwang and Seung Ki Baek for fruitful discussions and Cosimo Lupo for pointing out the similar feature in the spin glass clock models. This work was supported from the Basic Science Research Program through the National Research Foundation of Korea funded by the Ministry of Science, ICT \& Future Planning (NRF-2017R1D1A1B03034669). 
\end{acknowledgments}


\begin{thebibliography}{}

\bibitem{Berezinskii1971}
V. L. Berezinskii, Zh. Eksp. Teor. Fiz. \textbf{59}, 907 (1971)
[Sov. Phys. JETP \textbf{32}, 493 (1971)].

\bibitem{KT1972}
J. M. Kosterlitz and D. Thouless, J. Phys. C \textbf{5}, L124 (1972).

\bibitem{BKTreview}
{\it 40 Years of Berezinskii-Kosterlitz-Thouless Theory}, edited by J. V. Jos\'e (World Scientific, London, 2013).

\bibitem{KT1973}
J. M. Kosterlitz and D. J. Thouless, J. Phys. C \textbf{6}, 1181 (1973).

\bibitem{KT1974}
J. M. Kosterlitz and D. J. Thouless, J. Phys. C \textbf{7}, 1046 (1974).

\bibitem{Jose1977}
J. V. Jos\'e, L. P. Kadanoff, S. Kirkpatrick, and D. R. Nelson, 
Phys. Rev. B \textbf{16}, 1217 (1977).

\bibitem{Kenna2005}
R. Kenna, arXiv:cond-mat/0512356; 
MCFA Annals, Vol. IV, \url{http://www.mariecurie.org/annals/}.

\bibitem{Elitzur1979}
S. Elitzur, R. B. Pearson, and J. Shigemitsu, 
Phys. Rev. D \textbf{19}, 3698 (1979).

\bibitem{Cardy1980}
J. L. Cardy, J. Phys. A: Math. Gen. \textbf{13}, 1507 (1980).

\bibitem{Frohlich1981}
J. Fr\"ohlich and T. Spencer, Comm. Math. Phys. \textbf{81}, 527 (1981).

\bibitem{Ortiz2012}
G. Ortiz, E. Cobanera, and Z. Nussinov, Nucl. Phys. B \textbf{854}, 780 (2012).

\bibitem{Borisenko2011}
O. Borisenko, G. Cortese, R. Fiore, M. Gravina, and A. Papa, 
Phys. Rev. E {\bf 83}, 041120 (2011).

\bibitem{Einhorn1980}
M. B. Einhorn, R. Savit, and E. Rabinovici, 
Nucl. Phys. B \textbf{170}, 16 (1980).

\bibitem{Hamer1980}
C. J. Hamer and J. B. Kogut, Phys. Rev. B \textbf{22}, 3378 (1980).

\bibitem{Nienhuis1984}
B. Nienhuis, J. Stat. Phys. \textbf{34}, 731 (1984).

\bibitem{Tobochnik1982}
J. Tobochnik, Phys. Rev. B \textbf{26}, 6201 (1982); \textbf{27}, 6972 (1983).

\bibitem{Challa1986}
M. S. S. Challa and D. P. Landau, Phys. Rev. B \textbf{33}, 437 (1986).

\bibitem{Tomita2002}
Y. Tomita and Y. Okabe, Phys. Rev. B {\bf 65}, 184405 (2002).

\bibitem{Baek2010b}
S. K. Baek, P. Minnhagen, Phys. Rev. E {\bf 82}, 031102 (2010).

\bibitem{Baek2013}
S. K. Baek, H. M\"akel\"a, P. Minnhagen, and B. J. Kim, 
Phys. Rev. E {\bf 88}, 012125 (2013).

\bibitem{Nelson1977}
D. R. Nelson and J. M. Kosterlitz, Phys. Rev. Lett. \textbf{39}, 1201 (1977).

\bibitem{Minnhagen1981}
P. Minnhagen and G. G. Warren, Phys. Rev. B \textbf{24}, 2526 (1981).

\bibitem{Baek2010a}
S. K. Baek, P. Minnhagen, and B. J. Kim, Phys. Rev. E {\bf 81}, 063101 (2010).

\bibitem{Lapilli2006}
C. M. Lapilli, P. Pfeifer, and C. Wexler, Phys. Rev. Lett. {\bf 96}, 140603 (2006).

\bibitem{Kumano2013}
Y. Kumano, K. Hukushima, Y. Tomita, and M. Oshikawa, 
Phys. Rev. B {\bf 88}, 104427 (2013).

\bibitem{Chatelain2014}
C. Chatelain, J. Stat. Mech. (2014) P11022.

\bibitem{Hwang2009}
C.-O. Hwang, Phys. Rev. E {\bf 80}, 042103 (2009).

\bibitem{WL1}
F. Wang and D. P. Landau, Phys. Rev. Lett. \textbf{86}, 2050 (2001).

\bibitem{WL2}
F. Wang and D. P. Landau, Phys. Rev. E \textbf{64}, 056101 (2001).

\bibitem{YangLeeZero}
C. N. Yang and T. D. Lee, Phys. Rev. {\bf 87} 404 (1952).

\bibitem{FisherZero}
M. E. Fisher, in \textit{Lectures in Theoretical Physics}, Vol. 7C, ed. W. E. Brittin (University of Colorado Press, Boulder, 1965), Chap.~1.

\bibitem{Peng2015}
X. Peng, H. Zhou, B.-B. Wei, J. Cui, J. Du, and R.-B. Liu, 
Phys. Rev. Lett. \textbf{114}, 010601 (2015).

\bibitem{Brandner2016}
K. Brandner, V. F. Maisi, J. P. Pekola, J. P. Garrahan, and C. Flindt,
Phys. Rev. Lett. \textbf{118}, 180601 (2017).

\bibitem{Janke2001}
W. Janke and R. Kenna, J. Stat. Phys. {\bf 102}, 1211 (2001). 

\bibitem{Denbleyker2014}
A. Denbleyker, Y. Liu, Y. Meurice, M. P. Qin, T. Xiang, Z. Y. Xie, J. F. Yu, and H. Zou,
Phys. Rev. D {\bf 89}, 016008 (2014); H. Zou, Ph.D. Thesis, University of Iowa, 2014.

\bibitem{Rocha2016}
J. C. S. Rocha, L. A. S. M\'{o}l, and B. V. Costa, Comp. Phys. Commun. \textbf{209}, 88 (2016).

\bibitem{Costa2017}
B. V. Costa, L. A. S. M\'{o}l, and J. C. S. Rocha, Comp. Phys. Commun. \textbf{216}, 77 (2017).

\bibitem{Rocha2014}
J. C. S. Rocha, S. Schnabel, D. P. Landau, and M. Bachmann,
Phys. Rev. E \textbf{90}, 022601 (2014).

\bibitem{Taylor2013}
M. P. Taylor, P. P. Aung, and W. Paul, Phys. Rev. E \textbf{88}, 012604 (2013).

\bibitem{Lee2010}
J. H. Lee, H. S. Song, J. M. Kim, and S.-Y. Kim, J. Stat. Mech. (2010) P03020.

\bibitem{footnote1}
The case for $p=12$ is written similarly as $\cos(2\pi n/p)$ is composed of $1/2$ and $\sqrt{3}/2$ but is not considered here because of unaffordable computational cost in the WL sampling.

\bibitem{Landau2004}
D. P. Landau, S.-H. Tsai, and M. Exler, Am. J. Phys. \textbf{72}, 1294 (2004).

\bibitem{Zhou2006}
C. Zhou, T. C. Schulthess, S. Torbr\"ugge, and D. P. Landau,
Phys. Rev. Lett. \textbf{96}, 120201 (2006).

\bibitem{Silva2006}
C. J. Silva, A. A. Caparica, and J. A. Plascak,
Phys. Rev. E \textbf{73}, 036702 (2006).

\bibitem{Tsai2007}
S.-H. Tsai, F. Wang, and D. P. Landau, Phys. Rev. E \textbf{75}, 061108 (2007).

\bibitem{Kwak2015}
W. Kwak, J. Jeong, J. Lee, and D.-H. Kim, Phys. Rev. E \textbf{92}, 022134 (2015).

\bibitem{Falcioni1982}
M. Falcioni, E. Marinari, M. L. Paciello, G. Parisi, and B. Taglienti, Phys. Lett. \textbf{108B}, 331 (1982).

\bibitem{Alves1992}
N. A. Alves, B. A. Berg, and S. Sanielevici, Nucl. Phys. B \textbf{376}, 218 (1992).

\bibitem{SM}
The numerical data of the leading zeros are tabulated in Table~\ref{tab:zero_upper} and Table~\ref{tab:zero_lower} for the upper and lower transitions, respectively.

\bibitem{Vogel2013}
T. Vogel, Y. W. Li, T. W\"ust, and D. P. Landau,
Phys. Rev. Lett. \textbf{110}, 210603 (2013).

\bibitem{Vogel2014}
T. Vogel, Y. W. Li, T. W\"ust, and D. P. Landau,
Phys. Rev. E \textbf{90}, 023302, (2014).

\bibitem{Valentim2015}
A. Valentim, J. C. S. Rocha, S.-H. Tsai, Y. W. Li, M. Eisenbach, C. E. Fiore, and D. P. Landau,
J. Phys.: Conf. Ser. \textbf{640}, 012006 (2015).

\bibitem{Ren2016}
Y. Ren, S. Eubank, and M. Nath, Phys. Rev. E \textbf{94}, 042125 (2016).

\bibitem{Chan2017}
C. H. Chan, G. Brown, and P. A. Rikvold, Phys. Rev. E \textbf{95}, 053302 (2017).

\bibitem{Lupo}
C. Lupo (private communication).

\bibitem{Lupo2017}
C. Lupo and F. Ricci-Tersenghi, Phys. Rev. B \textbf{95}, 054433 (2017).

\bibitem{Nobre1986}
F. D. Nobre and D. Sherrington, J. Phys. C: Solid State Phys. \textbf{19}, L181 (1986).

\bibitem{Ilker2013}
E. Ilker and A. N. Berker, Phys. Rev. E \textbf{87}, 032124 (2013).

\bibitem{Ilker2014}
E. Ilker and A. N. Berker, Phys. Rev. E \textbf{90}, 062112 (2014).

\end{thebibliography}
\end{document}